\documentclass[aps,prl,letterpaper,twocolumn,tightenlines,superscriptaddress,notitlepage,longbibliography]{revtex4-2}
\usepackage{qcircuit}
\usepackage[dvips]{graphicx}
\usepackage{amsmath,amssymb,amsthm,mathrsfs,amsfonts,dsfont}
\usepackage{mathrsfs}
\usepackage{mathtools}
\usepackage{subfigure, epsfig}
\usepackage{braket}
\usepackage{bm}
\usepackage{enumerate}
\usepackage{xcolor}
\usepackage{comment}
\usepackage{scalefnt}
\usepackage[colorlinks = true, 
citecolor = blue,
linkcolor = blue,
urlcolor = blue]{hyperref}

\usepackage{pagecolor}
\usepackage[T1]{fontenc}
\usepackage{lmodern}
\usepackage[utf8]{inputenc}
\usepackage{microtype}

\usepackage{tikzscale}
\usepackage{pgfplots}
\pgfplotsset{compat=newest}
\usetikzlibrary{plotmarks}
\usetikzlibrary{arrows.meta}
\usetikzlibrary{external}
\usepgfplotslibrary{patchplots}
\usepackage{grffile}
\usepackage{enumitem}
\usepackage{makecell}

\DeclareMathOperator{\Tr}{Tr}

\newcommand{\coleq}{\mathrel{\mathop:}\nobreak\mkern-1.2mu=}
\newcommand{\eqcol}{\mkern-1.2mu=\mathrel{\mathop:}\nobreak}

\newcommand{\mc}{\mathcal}

\newcommand{\mr}{\mathrm}
\newcommand{\mbb}{\mathbb}

\newcommand{\ketbra}[2]{{\vert #1 \rangle \langle #2 \vert}}

\newcommand{\Pn}{{\mathsf{P}^n}}
\newcommand{\pt}{{\mathrm{pt}}}

\newtheorem{theorem}{Theorem}
\newtheorem{proposition}[theorem]{Proposition}%

\definecolor{applegreen}{rgb}{0.55, 0.71, 0.0}

\newcommand{\comments}[1]{}
\usepackage{algorithm}  
\usepackage[noend]{algpseudocode}  
\newcommand{\algorithmfootnote}[2][\footnotesize]{%
  \let\old@algocf@finish\@algocf@finish%
  \def\@algocf@finish{\old@algocf@finish%
    \leavevmode\rlap{\begin{minipage}{\linewidth}
    #1#2
    \end{minipage}}%
  }%
}
\usepackage{xparse}
\NewDocumentCommand{\LeftComment}{s m}{%
  \Statex \IfBooleanF{#1}{\hspace*{\ALG@thistlm}}\(\triangleright\) #2}
\algnewcommand{\LineComment}[1]{\Statex // #1}

\begin{document}

\title{Enhancing quantum noise characterization via extra energy levels}

\author{Senrui Chen}
    \email{csenrui@gmail.com}
    \affiliation{Pritzker School of Molecular Engineering, The University of Chicago, Chicago, Illinois 60637, USA}
\author{Akel Hashim}
    \affiliation{Quantum Nanoelectronics Laboratory, Department of Physics, University of California at Berkeley, Berkeley, CA 94720, USA}
    \affiliation{Applied Math and Computational Research Division, Lawrence Berkeley National Lab, Berkeley, CA 94720, USA}
\author{Noah Goss}
    \affiliation{Quantum Nanoelectronics Laboratory, Department of Physics, University of California at Berkeley, Berkeley, CA 94720, USA}
    \affiliation{Applied Math and Computational Research Division, Lawrence Berkeley National Lab, Berkeley, CA 94720, USA}
\author{Alireza Seif}
    \thanks{Present address: IBM Thomas J. Watson Research Center, Yorktown Heights, NY 10598, USA}
    \affiliation{Pritzker School of Molecular Engineering, The University of Chicago, Chicago, Illinois 60637, USA}
\author{Irfan Siddiqi}
    \affiliation{Quantum Nanoelectronics Laboratory, Department of Physics, University of California at Berkeley, Berkeley, CA 94720, USA}
    \affiliation{Applied Math and Computational Research Division, Lawrence Berkeley National Lab, Berkeley, CA 94720, USA}
\author{Liang Jiang}
    \email{liangjiang@uchicago.edu}
    \affiliation{Pritzker School of Molecular Engineering, The University of Chicago, Chicago, Illinois 60637, USA}
\date{\today}

\begin{abstract}
Noise is a major challenge for building practical quantum computing systems. Precise characterization of quantum noise is crucial for developing effective error mitigation and correction schemes. However, state preparation and measurement (SPAM) errors on many current platforms can introduce large ambiguity into conventional noise characterization methods. In this work, we propose a scheme for enhancing quantum noise characterization using additional energy levels. 
We first develop a comprehensive theory on the identifiability of $n$-qudit SPAM noise given high-quality single-qudit control, showing the existence of gauge freedoms which can be completely described using subsystem depolarizing maps.
We then show how to use these extra energy levels to reduce the gauge ambiguity in characterizing both SPAM and gate noise in the qubit subspace.
We experimentally implement these ideas on a superconducting quantum computing device and demonstrate a qutrit-enabled enhancement in noise characterization precision.

\end{abstract}

\maketitle

\section{Introduction}

Quantum computing promises powerful computational capabilities~\cite{nielsen2002quantum}, but its practical realization is hindered by quantum noise.
As we transition from the Noisy Intermediate-Scale Quantum (NISQ) era to the early stages of fault-tolerant quantum computing (FTQC), the ability to accurately characterize noise is essential for developing effective error mitigation and correction techniques. State preparation and measurement (SPAM) errors, in particular, present a significant challenge in this regard, causing ambiguity in traditional noise characterization methods.

A key difficulty arises from the presence of gauge freedoms~\cite{nielsen2021gate} -- transformations in the noise model parameters that do not affect any observable outcomes. 
These gauge ambiguities hinder us from correctly identifying errors in different components of a quantum computer.
The simplest example is that noises in state preparation (SP) and measurement (M) are generally related by a gauge transformation, thus cannot be distinguished from each other using conventional methods. 
Such gauge ambiguity originating from SPAM noise can also couple to gate noises, causing non-identifiability there~\cite{chen2023learnability}.
Though such ambiguity does not affect our ability to predict or mitigate noisy dynamics~\cite{chen2025disambiguating}, it is highly desirable to separate SP and M errors and identify the true gauge to enhance the characterization of quantum processors.
Various methods have been proposed to address these gauge ambiguity, such as by suppressing SP noise via algorithmic cooling~\cite{laflamme2022algorithmic} and by introducing noiseless entangling gates~\cite{lin2021independent}. 
Most of these methods need to rely on idealized assumptions that may or may not be experimentally justified.

In this work, we propose a novel approach to enhance noise characterization by exploiting additional energy levels beyond the qubit subspace -- specifically, using qudit (i.e., d-level) systems. We first develop a theoretical framework for understanding SPAM noise identifiability in n-qudit systems under the assumption of high-fidelity single-qudit control. We show that the gauge freedoms can be fully described by subsystem depolarizing maps and quantify the corresponding gauge ambiguity. We also provide efficient protocols to identify all SPAM noise parameter up to the gauge degrees of freedom.

Building on this, we introduce a protocol for qutrit-enhanced SPAM characterization that effectively confines gauge freedoms using information from higher energy levels. This allows for more precise estimation of SPAM in the qubit subspace, which in turn enables more precise characterization of gate noise parameter in the paradigm of Pauli noise learning.
We experimentally implement this scheme on a superconducting quantum processor and demonstrate a significant reduction in gauge ambiguity. Our results demonstrate how the capability to precise control qudit systems can enhance applications such as quantum noise characterization.

\section{Single-qudit SPAM characterization}
    \medskip
    
    Let us start with the task of characterizing SPAM noise of a single qubit. 
    We will assume the ability to perform perfect single-qubit unitary control. This is a reasonable assumption for many existing experimental platforms where the single-qubit control imperfection is order-of-magnitude weaker than SPAM or multi-qubit gates~\cite{hashim2020randomized,van2023probabilistic,ferracin2024efficiently}.
    Suppose the system can be initialized to some noisy ground state 
    \begin{equation}
    \rho_0 = (1-\varepsilon^S_1)\ketbra{0}{0}+ \varepsilon^S_1\ketbra{1}{1},   
    \end{equation}
    and that a noisy computational basis measurement with the following POVM elements can be conducted
    \begin{equation}
    \begin{aligned}
      E_0 &= (1-\varepsilon_{1,0}^M)\ketbra{0}{0}+\varepsilon^M_{0,1}\ketbra{1}{1},\\
      E_1 &= \varepsilon^M_{1,0}\ketbra{0}{0}+(1-\varepsilon_{0,1}^M)\ketbra{1}{1}.
    \end{aligned} 
    \end{equation}
    Here, $\varepsilon_1^S$ is the residual population on $\ket{1}$ and $\varepsilon_{1,0}^M$ is the probability of outputting $\ket{1}$ when measuring $\ket{0}$, similar for $\varepsilon_{0,1}^M$. 
    Note that, this model only describes an ``incoherent'' SPAM noise, but can always be enforced by randomly applying Pauli $Z$ gates between the SPAM.
    The question is then to characterize $\bm\varepsilon\coleq[\varepsilon_1^S,\varepsilon_{1,0}^M,\varepsilon_{0,1}^M]$. We further assume each entry of $\bm\varepsilon$ is sufficiently close to $0$, and informally denote them to be of order $\varepsilon$.

    Consider the following experiments: Initialize the qubit to $\rho_0$, apply either Pauli $I$ or Pauli $X$ gate, and then measure with $\{E_l\}_{l=0,1}$. Use $\Pr(j;U)\coleq\Tr(E_j U\rho U^\dagger)$ to denote the probability of outputting $j\in\{0,1\}$ when the applied gate is $U\in\{I,X\}$. We have (recall that gates are assumed to be noiseless)
    \begin{equation}
        \begin{aligned}
            \Pr(1;I) &= \varepsilon_1^S + \varepsilon^M_{1,0}+ O(\varepsilon^2)
            =\bm\varepsilon\cdot[1,0,1] + O(\varepsilon^2),\\
            \Pr(0;X) &= \varepsilon_1^S + \varepsilon^M_{0,1}+ O(\varepsilon^2)=\bm\varepsilon\cdot[1,1,0] + O(\varepsilon^2),
        \end{aligned}
    \end{equation}
    which, up to the first order, give us two linearly-independent equations about $\bm\varepsilon$.
    This turns out to be the maximum number of linearly-independent equations we can obtain.
    To see this, consider the following depolarizing gauge transformation~\cite{chen2024efficient}: 
    \begin{equation}\label{eq:gauge}
        \rho\mapsto \rho'\coleq \mc D_p(\rho_0),~E_l\mapsto E'_l\coleq \mc D_p^{-1}(E_l),
    \end{equation}
    where $\mc D_p(\cdot)\coleq (1-p)(\cdot)+p\Tr(\cdot)I/2$ is the depolarizing map with some parameter $p$. 
    Since any single-qubit unitary operation commutes with $D_p$, we have $ \Tr(E_l' \mc U(\rho')) = \Tr(E_l\mc U(\rho))$, thus no experiments can observe such transformation.
    Up to the first order, the depolarizing gauge transformation acts as
    \begin{equation}
        \bm\varepsilon \mapsto \bm\varepsilon + p/2\times[1,-1,-1] + O(\varepsilon^2).
    \end{equation}
    Though no experiment can exactly resolve this degree of freedom (DOF), we can bound the ambiguity using the positivity of the noise model, i.e., each entry of $\bm\varepsilon$ must be non-negative. This gives, again up to the first order,
        $-\varepsilon^S_1\le p/2\le \min\{\varepsilon_{0,1}^M,\varepsilon_{1,0}^M\},$
    and that each entry of $\bm\varepsilon$ can only be determined up to an gauge ambiguity of 
    \begin{equation}
        A = \varepsilon^S_1 +  \min\{\varepsilon_{0,1}^M,\varepsilon_{1,0}^M\}.
    \end{equation} 
    Later we will see that $A$ also quantifies the ambiguity in characterizing multi-qubit gate noise. From the above expression, it is clear that one can reduce the gauge ambiguity by suppressing SPAM noise. For example, one can use algorithmic cooling (or, heralded state preparation) to suppress state preparation noise~\cite{laflamme2022algorithmic}. However, there is always a limit in suppressing qubit SPAM noise, which put constraints in precise noise characterization. 

    \bigskip

    Now we introduce a SPAM noise characterization protocol enhanced by extra energy levels. 
    For this purpose, we will first generalize the above procedure to characterize SPAM noise of a single qudit (i.e., a $d$-level system). We will assume the ability to perform noiseless single-qudit control, which is justified by the fact that high-fidelity single-qutrit/ququart control have been experimentally demonstrated~\cite{morvan2021qutrit,yurtalan2020implementation}. The noisy initial state and computational-basis measurement are modeled, respectively, by
    \begin{equation}
        \begin{aligned}
            \rho_0=\sum_{j=0}^{d-1}\varepsilon^S_{j}\ketbra{j}{j},\quad
            E_l=\sum_{k=0}^{d-1}\varepsilon^M_{l,k}\ketbra{k}{k},
        \end{aligned}
    \end{equation}
    which can be enforced by applying random phase gates between SPAM. Thanks to the normalization condition, we can choose the independent parameters to be $\bm\varepsilon = \{\varepsilon_j^S;\varepsilon_{l,k}^M|j\neq0,k\neq l\}$, consisting of $d^2-1$ parameters. Since all the entries of $\bm\varepsilon$ can be interpreted as some error probability, we assume all of them are sufficiently close to $0$, written as of order $\varepsilon$.

    Now we study how to learn $\bm\varepsilon$. As shown in the Supplemental Materials~\footnote{\label{ft:sm} See Supplemental Materials.}, using as few as $2(d-1)$ different generalized X gates, one can obtain $d^2-2$ linearly-independent equations about $\bm\varepsilon$, under a first-order approximation; The one degree freedom left behind is explained by the qudit depolarizing gauge, same as Eq.~\eqref{eq:gauge} but with the qudit depolarizing map defined as $\mc D_p(\cdot)\coleq (1-p)(\cdot)+p\Tr(\cdot)I/d$.
    This transforms the parameters in $\bm\varepsilon$ according to:
    \begin{equation}
        \varepsilon_j^S\mapsto\varepsilon_j^S+p/d +O(\varepsilon^2),\quad\varepsilon_{l,k}^M\mapsto \varepsilon_{l,k}^M - p/d +O(\varepsilon^2)
        . 
    \end{equation}
    Invoking the positivity constraint, 
    one has $-\min\{\varepsilon_j^S\}\le p/d\le\min\{\varepsilon^M_{l,k}\}$.
    The gauge ambiguity is then given by
    \begin{equation}
        A = \min_{j\neq 0}\{\varepsilon_j^S\} + \min_{k\neq l}\{\varepsilon^M_{l,k}\}.
    \end{equation}
    In other words, the gauge ambiguity is only determined by the minimum initialization error and the minimum measurement error. 
    Consider, for example, a superconducting qubit. Apart from the ground state and first excited state that are used as the qubit basis ($\ket{0},\ket{1}$), there are higher excited energy levels ($\ket{2},\cdots$). One can expect the initialized state to be a thermal state on all the energy levels, and the population on $\ket{2}$ is order-of-magnitude smaller than that on $\ket{1}$, i.e., $\varepsilon^S_2\ll\varepsilon^S_1$; 
    Besides, it is reasonable to expect the probability of outputting $\ket{2}$ to be smaller than $\ket{1}$ while measuring $\ket{0}$, i.e., $\varepsilon_{2,0}^M\le\varepsilon_{1,0}^M$, though this depends on the details of how measurement signal is processed and classified~\cite{krantz2019quantum}.
    Therefore, we expect the qutrit-level information to help yield a much smaller gauge ambiguity than the qubit-only protocol. Note that such qutrit-enhanced protocol is fundamentally different from the cooling type of method~\cite{laflamme2022algorithmic}, as it does not change the qubit SPAM noise rate; Instead, it couples extra energy levels into one gauge using qutrit control to have a stronger positivity constraint.
    
\begin{figure}[!htp]
        \centering
        \includegraphics[width=0.99\linewidth]{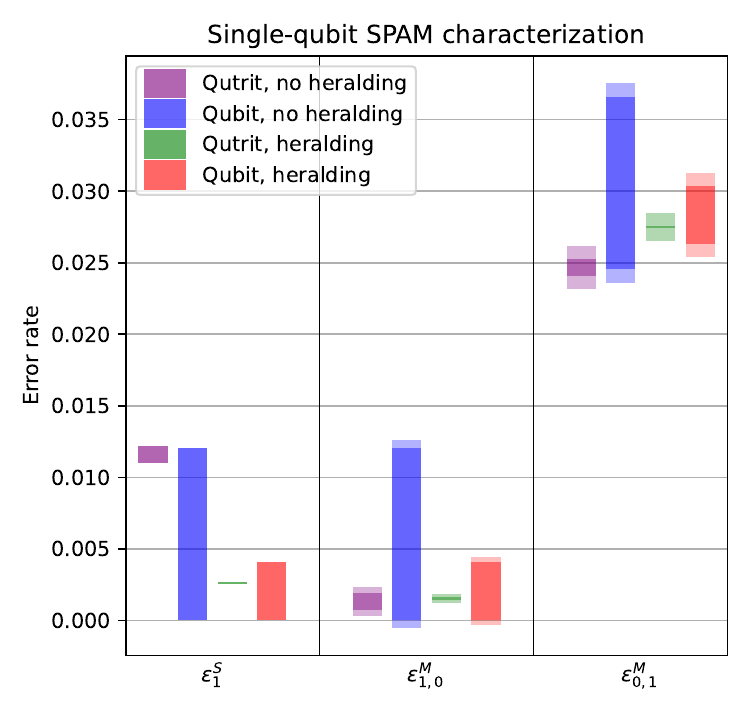}
        \caption{
        Comparison of qubit-only and qutrit-enhanced characterization of a single qubit, based on experimental data taken from a superconducting qutrit on 23-07-04. The solid bar represents the region of parameters obeying the positivity constraints. The light bar represents one standard error. Whether heralding is used or not, qutrit-enhanced method gives a much smaller gauge ambiguity than qubit-only method.}
        \label{fig:1q_spam_main}
    \end{figure}

    We experimentally demonstrate and compare the qutrit-enhanced and qubit-only SPAM characterization protocol. 
    Besides, we also combine them with (and without) a heralding initialization procedure: Measure the qudit right after initialization and discard the experiment if the output is not $\ket{0}$. This is similar to the algorithmic cooling procedure analyzed in Ref.~\cite{laflamme2022algorithmic} which can suppress the state preparation noise. From Fig.~\ref{fig:1q_spam_main}, one can see that, whether using heralding or not, the qutrit-enhanced protocol can significantly reduce the gauge ambiguity compared to the qubit-only protocol. 
    
    \section{Multi-qudit SPAM characterization}

    A practical quantum computing platform should have multiple qudits and support entangling operations. There can often be non-negligible correlations among the SPAM noise of different qudits~\cite{Bravyi2021Mitigating}, which motivates us to extend the above analysis to multi-qudit systems. Use $\ket{\bm j}$ for $\bm j \in\mbb Z_d^n$ to denote the computational basis state for an $n$-qudit system. We again assume the noisy initialization and computational-basis measurement are twirled into
    \begin{equation}\label{eq:multi_spam_model}
        \begin{aligned}
            \rho_0=\sum_{\bm j\in\mbb Z_d^n}\varepsilon^S_{\bm j}\ketbra{\bm j}{\bm j},\quad
            E_{\bm l}=\sum_{k\in\mbb Z_d^n}\varepsilon^M_{\bm l,\bm k}\ketbra{\bm k}{\bm k}.
        \end{aligned}
    \end{equation}
    This model can describe any correlated SPAM noise among the $n$ qudits. Thanks to normalization, we can choose the independent parameters to be $\bm\varepsilon = \{\varepsilon_{\bm j}^S;\varepsilon_{\bm l,\bm k}^M|\bm j\neq\bm 0,\bm k\neq \bm l\}$, consisting of $d^{2n}-1$ parameters that are sufficiently close to $0$. 
    We also assume the ability to perform any noiseless single-qudit unitary gate in parallel, but do not consider multi-qubit entangling gates for now.
    Note that, while noiseless entangling gates can indeed be used to resolved SPAM gauge ambiguity~\cite{lin2021independent}, noisy entangling gates cannot achieve so in general as they would introduce a new set of noisy parameters~\cite{chen2023learnability}.
    We have the following result on learning $\bm\varepsilon$ within this model, which might be of independent interest.
    
    \begin{proposition}\label{prop:quditSPAM_main}
    For an $n$-qudit system, given perfect single-qudit control, there are $2^n-1$ gauge degrees of freedom (DOFs) for incoherent SPAM noise, up to a first-order approximation; Furthermore, there exists a protocol using no more than $2d^n$ circuits to estimate all the learnable DOFs of the noise. 
    \end{proposition}
    Here, the gauge DOFs can be described by the subsystem depolarizing maps. Given a non-empty subset of qubits $\Omega\subseteq[n]$, the subsystem depolarizing map on $\Omega$ is defined as
    \begin{equation}\label{eq:sdg}
    \mc D_p^{\Omega}(\cdot) = (1-p)(\cdot) + p\Tr_\Omega(\cdot)\otimes I_{\Omega}/d^{|\Omega|}.
    \end{equation}
    By replacing $\mc D_p$ with $\mc D_p^{\Omega}$ in Eq.~\eqref{eq:gauge} we obtain a gauge transformation, as $\mc D_p^{\Omega}$ commutes with any parallel single-qudit unitaries. One can show each choice of $\Omega$ yields a linearly-independent transformation, thus gives $2^n-1$ gauge DOF in total. Each of them is controlled by an independent gauge parameter, and the gauge parameters need to satisfy the positivity constraints; On the other hand, we can construct no more than $2d^n$ circuits consisting solely of generalized Pauli $X$ gates to determine all the other $(d^{2n}-1)-(2^n-1)$ learnable DOFs, to the first order. 
    Here, a generalized $X$ gate is defined as $X_{j,k} =\ketbra{j}{k}+\ketbra{k}{j}$.
    More details are provided in SM~\cite{Note1}.

\begin{figure}
        \centering
        \includegraphics[width=\linewidth]{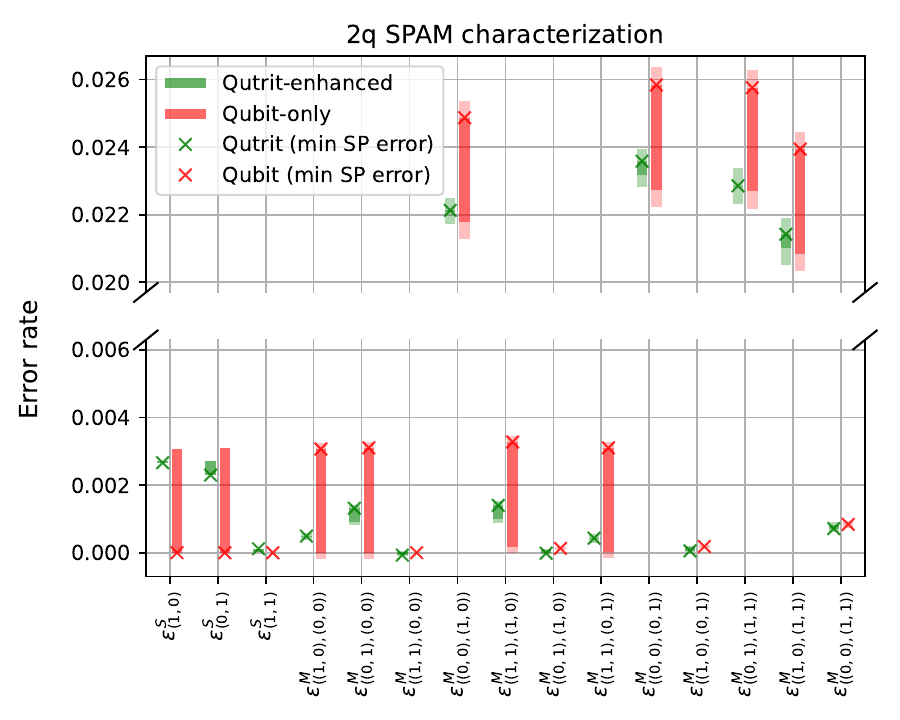}
        \caption{Comparison of qubit-only (Left) and qutrit-enhanced (Right) SPAM characterization of 2 qubits. Data were taken from superconducting qutrits on 24-07-11. Heralding initialization is used for both experiments. Solid bars represent the region of parameters obeying positivity constraints. 
        Light bars represents one standard error for deciding the region. The crosses denote one specific choice of gauge parameters that minimize the state-preparation error, which is used to highlight the solutions when the solid bars are too short to be seen.}
        \label{fig:2q_spam_main}
    \end{figure}

    We experimentally demonstrate these ideas with $2$ superconducting qubits and qutrits. In both case, there are $3$ gauge DOFs, including $2$ independent depolarizing gauges on each qudit and $1$ correlated depolarizing gauge, i.e., $\Omega=\{1\},\{2\},\{1,2\}$. 
    We are only interested in learning the SPAM noise parameters within the qubit subspace, but allows using the qutrit level to confine gauge parameters. Heralding initialization is turned on for both experiments.
    The results are shown in Fig.~\ref{fig:2q_spam_main}. One can clearly see that, while both protocols give consistent estimation, the qutrit-enhanced scheme significantly reduces the gauge ambiguity compared to the qubit-only scheme.

    \section{Application in Pauli Noise Characterization}

    We now explain how the aforementioned protocol can be applied to enhance noise characterization beyond SPAM. From now on, we will focus on an $n$-qubit system. A useful formalism is known as Pauli noise characterization~\cite{flammia2020efficient,wallman2016noise,erhard2019characterizing,hashim2020randomized,chen2022quantum,chen2024efficient}, which is applied to quantum error mitigation~\cite{van2023probabilistic,ferracin2024efficiently,kim2023evidence} and error correction~\cite{iyer2022efficient,chen2022calibrated,hashim2023benchmarking}, etc. Assuming the system support a set of multi-qubit entangling Clifford gate $\mc G$, the noisy implementation of each $\mc G$ is modeled as $\widetilde{\mc G} = \mc G\circ\Lambda_{\mc G}$ where $\Lambda_{\mc G}$ is a Pauli channel, satisfying the following forms
    \begin{equation}
        \Lambda_{\mc G}(\cdot)=\sum_{a\in \Pn}p^{\mc G}_aP_a(\cdot)P_a = \sum_{b\in\Pn}\lambda_b^{\mc G}P_b\Tr[P_b(\cdot)]/2^n,
    \end{equation}
    where $\{p^{\mc G}_a\}$ form a probability distribution known as the Pauli error rates~\cite{nielsen2002quantum}, and $\{\lambda_b^{\mc G}\}$ are known as the Pauli fidelities or Pauli eigenvalues~\cite{flammia2020efficient}. Noise on any single-qubit gates is either ignored or assumed gate-independent and can thus be absorbed into the multi-qubit gate layer. The SPAM are assumed to suffer from symmetrized Pauli noise channels whose Pauli eigenvalues only depend on the support of the Pauli but not distinguishing $X,Y,Z$ (denoted as $\Lambda^S$ and $\Lambda^M$). We remark that all of the above assumptions can be enforced by Randomized compiling, a standard technique to convert a generic noise into a Pauli noise model, given sufficiently good single-qubit control~\cite{wallman2016noise,hashim2020randomized}. 

    The gauge ambiguity of SPAM couples with the quantum gates and leads to ambiguity in identifying gate-dependent Pauli noise~\cite{chen2023learnability,chen2024efficient,huang2022foundations}. For example, it is shown in Ref.~\cite{chen2023learnability} that $\lambda_a^{\mc G}$ is gauge-independently (or, SPAM-independently) identifiable if and only if the ideal Clifford gate $\mc G$ does not change the support of $P_a$. Take $n=2$ and $\mc G = \mr{CZ}$ for example (omitting the superscript of $\mc G$): while $\lambda_{ZI}$ and $\lambda_{XX}$ are identifiable, $\lambda_{XI}$ is not. This is because such non-identifiable eigenvalues can be changed under a subsystem depolarizing gauge (see Eq.~\eqref{eq:sdg}).
    Nevertheless, using the positivity constraints, one can still bound the non-identifiable eigenvalues up to certain gauge ambiguity. 

    Concretely, consider characterizing $\mc G=\mr{CZ}$ on a $2$-qubit system. By preparing an $+1$ eigenstate of $P_a$, consecutively applying $2t+1$ noisy CZ (with randomized compiling), and measuring the expectation value of $\mc G(P_a)$ yields
    \begin{equation}
        F_a(t) = \lambda^S_a\lambda^M_{\mc G(a)}\lambda_a(\lambda_{a}\lambda_{\mc G(a)})^t.
    \end{equation}
    By estimating $F_a$ at different depth $t$ and fit to an exponential decay model, one can obtain consistent estimators for the intercept $\lambda^S_a\lambda^M_{\mc G(a)}\lambda_a$ and decay rate $\lambda_{a}\lambda_{\mc G(a)}$. 
    Meanwhile, we can use the SPAM characterization protocol described earlier to estimate $\lambda_a^S\lambda_{\mc G(a)}^M$ and divide it from the intercept estimator to estimate $\lambda_a$.
    This procedure is called ``intercept cycle benchmarking'' in Ref.~\cite{chen2023learnability} and is also explored in Ref.~\cite{van2023probabilistic,van2024techniques,carignan2023error}.
    Note that when $\mc G$ preserves the support of $P_a$, $\lambda^S_a\lambda^M_{\mc G(a)}$ is identifiable to the first order and so is $\lambda_a$, consistent with the fact that it is identifiable; When $\mc G$ changes the support of $P_a$, $\lambda^S_a\lambda^M_{\mc G(a)}$ suffers from gauge ambiguity, which then translates into the ambiguity in estimating $\lambda_a$. Consequently, the smaller the SPAM gauge ambiguity is, the more precise $\lambda_a$ can be bounded. That is why we expect qutrit-enhanced SPAM characterization protocol to also enhance gate noise characterization.

    \begin{figure}[t]
        \centering
        \includegraphics[width=\linewidth]{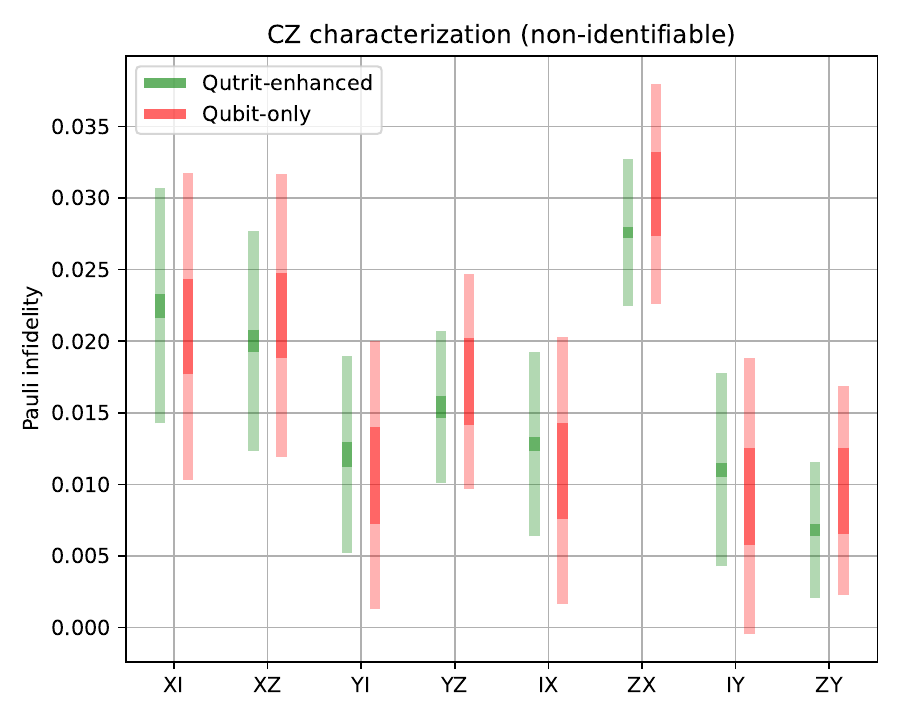}
        \caption{Experimentally bounding the non-identifiable Pauli fidelities for CZ using data from both qubit-only and qutrit-enhanced SPAM characterization scheme. Heralded initialization is used in all experiments. Same dataset as in Fig.~\ref{fig:2q_spam_main}. The light bar represents one standard error.}
        \label{fig:gate_main}
    \end{figure}
    
\begin{figure}[t]
        \centering
        \includegraphics[width=\linewidth]{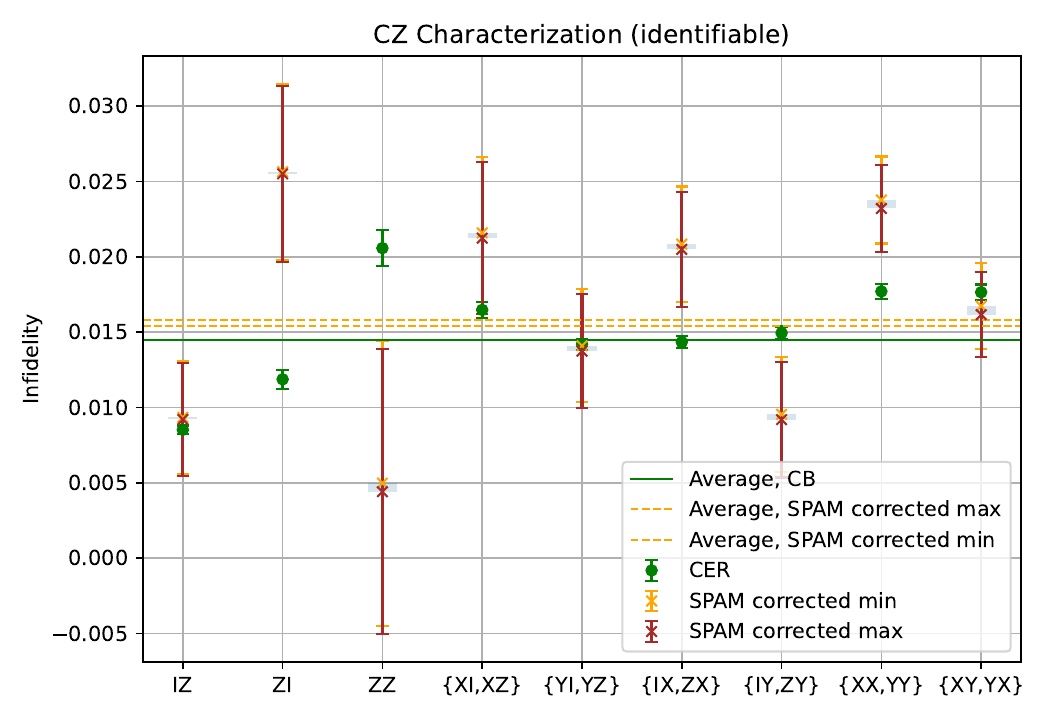}
        \caption{Comparison of Cycle Error Reconstruction (CER) and our SPAM-corrected protocol on the identifiable eigenvalues. Error bars represent one standard error. The horizontal lines show the average fidelity predicted by both schemes -- CB denotes average fidelity from CER.
        The SPAM corrected max/min refer to two extreme choices of the gauge parameters in our protocol.
        The blue shaded region denotes the difference in estimates from the two choices of gauge, which is due to the first-order approximation used in our methods. 
        }
        \label{fig:gate_verification}
    \end{figure}

    Fig.~\ref{fig:gate_main} shows the experimental results of bounding non-identifiable Pauli eigenvalues for CZ, using SPAM characterization data from both qubit-only and qutrit-enhanced scheme. Clearly, for all those eigenvalues, the latter scheme gives a smaller gauge ambiguity than the former, demonstrating an enhancement in estimation precision. 
    As a sanity check, we also implement the standard cycle error reconstruction (CER) protocol~\cite{erhard2019characterizing,carignan2023error} to estimate the identifiable combinations of Pauli eigenvalues (e.g., $\sqrt{\lambda_{XI}\lambda_{XZ}}$) and compare with the estimates from our protocol. 
    The results are shown in Fig.~\ref{fig:gate_verification}. For most eigenvalues or pairs of eigenvalues, the estimates from both schemes match within around one standard error, supporting the validity of our theory. For $ZI$ and $ZZ$ eigenvalues, the discrepancy is more significant but still within $2$ standard errors. We expect such difference to either come from out-of-model errors such as non-Markovian and non-time-stationary sources of noise or statistical fluctuation.

\section{Discussion}

In this work, we present a comprehensive theory on the gauge ambiguity and characterization of multi-qudit SPAM noise, given noiseless single-qudit control. As a concrete application, we show how qutrit-enhanced protocol offers improved characterization for qubit-subspace SPAM noise and gate-associated Pauli noise, which is supported by experimental results. 
Our results can be viewed as an extension of the recent self-consistent Pauli noise learning framework~\cite{chen2023learnability,chen2024efficient} to qudit systems.

Many open questions and future directions remain. One immediate extension is to account for imperfections in single-qutrit control. Although noiseless (or with gate-independent noise) single-qubit gate is a standard assumption that has been widely validated in experiments~\cite{hashim2020randomized,van2023probabilistic,kim2023evidence}, single-qudit gate might be more noisy due to their complexity. 
Another source of imperfection not considered here is non-Markovian noise. In particular, a form of non-Markovian noise in our experiments can be leakage and seepage between the third energy level and the qubit subspace, which may partly explain the discrepancy seen in our experimental results. 
Extending our protocol to handle imperfect qudit control and non-Markovian noise remains an important challenge.

Another promising direction involves scaling to larger systems. In our theory of $n$-qudit SPAM noise characterization, we allow the most general noise channels that contain exponentially many parameters. In practice, one can impose a quasi-local noise model to make the number of parameters tractable while capture most of the underlying noise. Many such examples exist in the literature of Pauli noise learning~\cite{van2023probabilistic,chen2024efficient,flammia2020efficient,wagner2023learning}. We believe such quasi-local assumptions can be incorporated into our multi-qudit SPAM noise characterization framework.

Finally, we emphasize that our protocol is complementary to cooling-based methods such as heralded state preparation~\cite{laflamme2022algorithmic}. Rather than reducing SPAM error rates themselves, our approach couples additional energy levels into a unified gauge framework, thereby enhancing identifiability through stronger positivity constraints. This distinction highlights the utility of qutrit control not only for quantum computation but also as a tool for more precise diagnostics of quantum noise.

\medskip
\noindent \emph{Note added.} While completing this project, we became aware of an independent and concurrent work by C. Haupt \emph{et al.}~\cite{Haupt2025}, which contains similar ideas of leveraging extra energy levels to characterize SPAM noise.

\bigskip

\begin{acknowledgments}
We thank Yunchao Liu, Zhihan Zhang, Ming Yuan, Long Nguyen, Larry Chen, Pranav Gokhale for helpful discussions. AS acknowledges support from a Chicago Prize Postdoctoral Fellowship in Theoretical Quantum Science. AS contributions were completed while he was affiliated with the University of Chicago. S.C. and L.J. acknowledge support from the ARO (W911NF-23-1-0077), ARO MURI (W911NF-21-1-0325), AFOSR MURI (FA9550-21-1-0209, FA9550-23-1-0338), DARPA (HR0011-24-9-0359, HR0011-24-9-0361), NSF (ERC-1941583, OMA-2137642, OSI-2326767, CCF-2312755, OSI-2426975), and the Packard Foundation (2020-71479). This material was funded in part by the U.S. Department of Energy, Office of Science, Office of Advanced Scientific Computing Research Quantum Testbed Program under Contract No. DE-AC02-05CH11231. N.G. acknowledges support from the Kavli ENSI Graduate Student Fellowship program.
\end{acknowledgments}

\bibliography{Qutrit.bib}

\newpage
\widetext
\begin{center}
\textbf{\large Supplemental Materials:\\ Enhancing quantum noise characterization via extra energy levels}
\end{center}
\setcounter{equation}{0}
\setcounter{figure}{0}
\setcounter{table}{0}
\setcounter{theorem}{0}
\setcounter{section}{0}
\makeatletter
\renewcommand{\thesection}{S\arabic{section}}
\renewcommand{\theequation}{S\arabic{equation}}
\renewcommand{\thefigure}{S\arabic{figure}}

\section{Theory of multi-qudit SPAM characterization}

\subsection{Notations}

We assume familiarity with standard quantum information notations, see e.g.~\cite{nielsen2002quantum}. A qudit is a $d$-dimensional quantum system, and an $n$-qudit system is the tensor product of $n$ qudits, with dimension $d^n$. The computational basis for an $n$-qudit system is denoted as $\{\ket{\bm j}\}$ for $\bm j\in\{0,\cdots,d-1\}^n$. We use the notation of $\{0,\cdots,d-1\}^n$ and $\mbb Z_d^n$ interchangably. Summation of a vector $\bm j$ is by default over $\mbb Z_d^n$ unless otherwise specified.

Define the indicator function $\mathds 1[\text{``statement''}]$ whose value is $1$ if the ``statement'' is true and $0$ elsewise. Define the Kronecker Delta function $\delta_{a,b}\coleq\mathds 1[a=b]$.

\subsection{Introduction and Setup}

In this section, we develop a theory on the characterization of $n$-qudit SPAM given perfect single-qudit control. 
Assume the system of interest has a single implementation of ground state initialization and computational basis readout, modelled by a density matrix $\rho_{0}$ and a POVM measurement $\{E_{\bm l}\}_{{\bm l}\in\mbb Z_d^n}$, respectively. 
We further assume both of them are incoherent, i.e., diagonal in the computation basis. 
We remark that this condition can be imposed by applying random phases after initialization and before readout, consistent with randomized compiling-based methods~\cite{wallman2016noise}.
In the language of gate set tomography, the model we study can be viewed as the following parameterized gate set~\cite{nielsen2021gate},
\begin{itemize}
    \item Noisy initialization: $\rho_0=\sum_{\bm j}c_{\bm j}\ketbra{\bm j}{\bm j}$.
    \item Noisy measurement: $\{E_{\bm l}=\sum_{\bm k}s_{\bm l,\bm k} \ketbra{\bm k}{\bm k}\}_{\bm l}$.
    \item Noiseless single-qudit gates: $\bigotimes_{i=1}^n U_i,~\forall U_i\in SU(d)$.
\end{itemize}
The model parameters to be extracted are $c_{\bm j}$ and $s_{\bm l,\bm k}$. 
The former can be interpreted as the initial population at level $\ket{\bm j}$. The latter can be interpreted as the confusion probability from $\ket{\bm k}$ to $\ket{\bm l}$ at the readout.
By normalization conditions, we have
\begin{equation}
    \sum_{\bm j} c_{\bm j} = 1;\quad \sum_{\bm l}s_{\bm l,\bm k} = 1,~\forall\bm k\in\mbb Z_{d}^{n}.
\end{equation}
Therefore, there are $(d^n-1) + d^n(d^n-1) = d^{2n}-1$ independent degrees of freedom. 
However, not all of these parameters are identifiable, as there exist gauge transformations that modify the model parameters but cannot be observed by any experiments within the gate set. Note that the gauge transformation must preserve all model assumptions, including perfect single-qudit gates and incoherent SPAM. In the following, we will focus on the following two questions
\begin{enumerate}
    \item What and how many degrees of freedom are identifiable, (or, ``learnable'')?
    \item How many experiments are needed to extract all learnable information?
\end{enumerate}
Our answers are summarized in the following,
\begin{proposition}\label{prop:quditSPAM}
    For an $n$-qudit system, given perfect single-qudit control, there are $2^n-1$ gauge degrees of freedom for incoherent SPAM noise, up to a first-order approximation. Furthermore, there exists a protocol using no more than $2d^n$ circuits to estimate all the learnable degrees of freedom of the noise.
\end{proposition}

\bigskip
We start by understanding the gauge degrees of freedom of this gate set.
Any experiment within the gate set can be specified by a list of $n$ single-qudit unitary $\{U_i\}_{i=1}^n$. (as the concatenation of any layer of noiseless single-qudit is just another layer of single-qudit gate). 
Define $\bm U\coleq \bigotimes_{i=1}^n U_i$.
The outcome distribution can be written as
\begin{equation}\label{eq:observable}
    \Pr(\bm l;\{U_i\}_{i=1}^n) = \Tr[E_{\bm l} \bm U\rho_0 \bm U^\dagger] = \sum_{\bm k,\bm j}c_{\bm j}s_{\bm l,\bm k}\prod_{i=1}^n|\bra{\bm k_i}U_i\ket{\bm j_i}|^2\eqcol\sum_{\bm k,\bm j}c_{\bm j}s_{\bm l,\bm k}u_{\bm k,\bm j}.
\end{equation}
Define the depolarizing map on a $k$-qudit system,
\begin{equation}
    \Lambda_p(\rho) = (1-p)\rho + p\frac{1}{d^k}I,
\end{equation}
where $I$ is the identity operator on the corresponding system. $p$ is a real number close to $0$. 
We will use two properties of $\Lambda_p$. First, it is self-conjugate, i.e., $\Lambda_p^+ = \Lambda_p$, where the conjugate is defined such that
\begin{equation}
    \Tr[E\Lambda(\rho)] = \Tr[\Lambda^\dagger(E) \rho]
\end{equation}
holds for all $E,\rho$.
Second, the inverse map of a depolarizing map is still a depolarizing map.
Indeed, let $p^*\coleq -p/(1-p)$, one can verify that
\begin{equation}
    \Lambda_{p^*}\circ\Lambda_p = \Lambda_p\circ\Lambda_{p^*} = \mathds 1.
\end{equation}
We thus have $\Lambda_p^{-1}=\Lambda_{p^*}$. 
Let $\Omega$ be a subset of $\{1,\cdots,d\}$, we use $\Lambda_{\Omega,p}$ to denote the depolarizing channel acting on the set of qudits indexed by $\Omega$ and trivially elsewhere. We now claim that the following is a valid gauge transformation for all non-empty $\Omega$,
\begin{equation}
    \rho_0\mapsto\Lambda_{\Omega,p}(\rho_0),
    \quad E_{\bm l}\mapsto \Lambda_{\Omega,p}^{-1}(E_{\bm l}),
    \quad \bigotimes_{i=1}^n\mc U\mapsto\Lambda_{\Omega,p}\circ\bigotimes_{i=1}^n\mc U\circ\Lambda_{\Omega,p}^{-1}.
\end{equation}
This transformation obviously does not change any observable distribution as in Eq.~\eqref{eq:observable}. We just need to verify it preserves our noise assumptions. It is easy to check the SPAM is still incoherent. To see the single-qudit unitary is still noiseless, note that the depolarizing map commute with any unitary channel acting on the same system. That is,
\begin{equation}
    \Lambda_p\circ\mc U(\rho) = (1-p)U\rho U^\dagger + p \frac{1}{d^{|\Omega|}}I = \mc U\circ\Lambda_p(\rho),\quad \forall \rho.
\end{equation}
Consequently, the single-qubit gate layer is preserved. We conclude this is a valid gauge transformation. By choosing different $\Omega$, we can construct $2^n-1$ different gauge transformations. We call them the \emph{subspace depolarizing gauges}.

\subsection{A first-order gauge theory}

In the following, we introduce a practically motivated assumption: The SPAM noise is reasonably small. That is, $\rho_0$ is close to the ideal ground state $\ketbra{\bm 0}{\bm 0}$ and $\{E_{\bm l}\}$ is close to the ideal computational basis measurement $\{\ketbra{\bm l}{\bm l}\}$. In terms of the model parameters, $c_{\bm 0}$ and $s_{\bm l,\bm l}$ for all $\bm l$ are close to $1$. (let $1-c_{\bm 0},1-s_{\bm l,\bm l}$ be of order $\varepsilon$.)
Without loss of generality, choose the $d^{2n}-1$ independent model parameters to be 
$$c_{\bm j},\forall\bm j\neq\bm 0\quad\text{and}\quad s_{\bm k,\bm l}, \forall\bm k\neq\bm l.$$ 
All these parameters are small values. In the following, our analysis will be based on first-order approximation.
Specifically, we will focus on linear functions of these functions, as a general function can always be expanded to the first order. 

We can all the independent noise parameters in a vector $\bm v\in\mbb R^{d^{2n}-1}$. Any linear function can also be represented by a vector $\bm f\in\mbb R^{d^{2n}-1}$, whose value evaluated at $\bm v$ is given by the standard inner product $\bm f\cdot \bm v$. A linear transformation on the noise parameters can also be represented by a vector $\bm t\in \mbb R^{d^{2n}-1}$ via $\bm v\mapsto\bm v + \bm t$.  
We call a function $\bm f$ to be first-order learnable if it is orthogonal to any gauge transformation $\bm g$, \textit{i.e.}, $\bm f\cdot \bm g = O(\varepsilon^2)$. 

Let us first check how the subspace depolarizing gauge transform the noise parameters. With some calculation, one can verify that
\begin{equation}
    \begin{aligned}
        c_{\bm j}&\mapsto (1-p)c_{\bm j} + \frac{p}{d^{|\Omega|}}
        \sum_{\substack{\bm i~s.t.\\ \bm i_{\neg\Omega}=\bm j_{\neg\Omega}}}c_{\bm i},\\
        s_{\bm l,\bm k}&\mapsto \frac{1}{1-p}s_{\bm l,\bm k} - \frac{p}{1-p}\frac{1}{d^{|\Omega|}}
        \sum_{\substack{\bm i~s.t.\\ \bm i_{\neg\Omega}=\bm j_{\neg\Omega}}}
        s_{\bm l,\bm i}.
    \end{aligned}
\end{equation}
Here, the subscript of $\neg\Omega$ selects the entries whose indexes are not in $\Omega$.
Take $p$ to be of the order of $\varepsilon$, we get
\begin{equation}
    \begin{aligned}
        c_{\bm j}&\mapsto\left\{\begin{aligned}
            &c_{\bm j} + \frac{p}{d^{|\Omega|}} + O(\varepsilon^2),&&\quad\text{if}~\bm j_{\neg\Omega}=\bm 0_{\neg\Omega}.\\
            &c_{\bm j} + O(\varepsilon^2),&&\quad\text{elsewise}.
        \end{aligned}\right.\\
        s_{\bm l,\bm k}&\mapsto\left\{\begin{aligned}
            &s_{\bm l,\bm k} - \frac{p}{d^{|\Omega|}} + O(\varepsilon^2),&&\quad\text{if}~{\bm k_{\neg\Omega}=\bm l_{\neg\Omega}}.\\
            &s_{\bm l,\bm k} + O(\varepsilon^2),&&\quad\text{elsewise}.
        \end{aligned}\right.\\
    \end{aligned}
\end{equation}
Ignoring the second-order term, the gauge transformation becomes linear. Concretely, the normalized linearized gauge transformation associated with $\Omega$ can be expressed as
\begin{equation}
    \bm{\widetilde{g}}_\Omega:~ c_{\bm j}\mapsto c_{\bm j} + \delta_{\bm j_{\neg\Omega},\bm 0_{\neg\Omega}},\quad s_{\bm l,\bm k}\mapsto s_{\bm l,\bm k} - \delta_{\bm k_{\neg\Omega},\bm l_{\neg\Omega}}.
\end{equation}
One can show that $\{\bm {\widetilde g}_\Omega\}$ for all non-empty $\Omega$ are linearly-independent. To see this, it would be easier to change to another basis. Define the mapping $\pt:\mbb Z_d^n\mapsto\{0,1\}^n$ according to $\pt(\bm j)_i = \mathds 1[\bm j_i\ne 0]$, and define the following transformations
\begin{equation}
    \bm{g}_\Omega:~ c_{\bm j}\mapsto c_{\bm j} + \delta_{\pt(\bm j),\Omega},\quad s_{\bm l,\bm k}\mapsto s_{\bm l,\bm k} - \delta_{\pt(\bm k-\bm l),\Omega}.
\end{equation}
Here, we slightly abuse the notation of $\Omega$ to represent both a subset of $\{1,\cdots,n\}$ and an $n$-bit string. There is an natural way to translate between these two representations: including the $i$th element if and only if the $i$th bit is $1$. $\{\bm g_\Omega\}$ are obviously linearly independent as they act non-trivially on disjoint set of $c_{\bm j}$ for different $\Omega$. On the other hand, they can be linearly represented by $\{\bm{\widetilde{g}}_\Omega\}$ via
\begin{equation}
    \bm g_\Omega = \sum_{\Omega'\subseteq\Omega,~\Omega'\neq\varnothing} (-1)^{|\Omega|-|\Omega'|}\bm{\widetilde{g}}_{\Omega'},
\end{equation}
according to the inclusion-exclusion principle. We conclude that $\{\bm{\widetilde g}_\Omega\}$ are linearly-independent. and thus give us $2^n-1$ gauge degrees of freedom.

Now we show that they are actually all the possible gauge transformations.
To see this, we show that by introducing $2^n-1$ additional linear constraints, all the noise parameters can be determined.
Specifically, by fixing the value of $c_{\Omega}$ for all $\Omega\in\{0,1\}^n$, $\Omega\neq\bm 0$ (called \emph{gauge parameters}), we now construct a protocol to determine all the other parameters. 
Define the single-qudit generalized Pauli X gates as
\begin{equation}
    X_\pi \coleq \sum_{j}\ketbra{\pi(j)}{j},\quad\forall \pi\in S_d.
\end{equation}
Let $\bm\pi=[\pi_1,\cdots,\pi_n]$ denote a vector of permutation. The action of $\bm\pi$ on $\bm j\in\mbb Z_d^n$ is entrywise. Denote $X_{\bm\pi}=X_{\pi_1}\otimes\cdots\otimes X_{\pi_n}$. An experiment using $X_{\bm\pi}$ yields the following probability distribution
\begin{equation}
    \Pr(\bm l; X_{\bm\pi}) = \sum_{\bm j}c_{\bm j}s_{\bm l,\bm{\pi}(\bm j)},\quad\forall \bm l\in\mbb Z_d^n.
\end{equation}
This gives us at most $2^n-1$ independent equations because of normalization. 
Let $\bm t\coleq \bm\pi^{-1}(\bm l)$.
Without loss of generality, focus on $\bm t\ne\bm 0$. We have
\begin{equation}\label{eq:perm_observable}
    \Pr(\bm \pi(\bm t);X_{\bm \pi}) = \sum_{\bm j}c_{\bm j}s_{\bm \pi(\bm t),\bm\pi(\bm j)} = c_{\bm t} + s_{\bm\pi(\bm t), \bm\pi(\bm 0)} + O(\varepsilon^2).
\end{equation}
Now we specify the design of $\bm\pi$. First, for all $\bm h\in\{1,\cdots,d-1\}^n$, choose $\bm\pi^{(\bm h)}$ such that
\begin{equation}
    \pi^{(\bm h)}_i = \left\{\begin{aligned}
        &\mathds 1,&&\quad\text{if~}h_i=1,\\
        &(1,h_i),&&\quad\text{if~}h_i>1.
    \end{aligned}\right.
\end{equation}
These experiments allow us to determine $c_{\bm j}$ and $s_{\bm j,\bm 0}$ for all $\bm j\ne\bm 0$. 
To see this, first look at the probability of seeing $\bm j$ in the experiment associated with identity (\textit{i.e.,} $\bm h = 1^n$), given by Eq.~\eqref{eq:perm_observable},
\begin{equation}\label{eq:prob_I}
    \Pr(\bm j;\bm I)= c_{\bm j}+s_{\bm j,\bm 0} + O(\varepsilon^2).
\end{equation}
For any $\bm j\in\{0,1\}^n\backslash\{\bm 0\}$, $c_{\bm j}$ is a gauge parameter, and the above equation help us determine $s_{\bm j,\bm 0}$ up to $O(\varepsilon^2)$; For $\bm j\notin\{0,1\}^n$, it is not hard to see there must exist an $\bm\pi^{(\bm h)}$ such that $\bm\pi^{(\bm h)}(\pt(\bm j)) = \bm j$. 
Again by Eq.~\eqref{eq:perm_observable}, we have the probability of seeing $\bm j$ in that experiment,
\begin{equation}\label{eq:prob_more}
    \begin{aligned}
        \Pr(\bm j;X_{\bm\pi^{(\bm h)}})&= c_{\pt(\bm j)} + s_{\bm j,\bm 0} + O(\varepsilon^2).
    \end{aligned}
\end{equation}
Combining Eqs.~\eqref{eq:prob_I} and~\eqref{eq:prob_more}, we can determine the values of $c_{\bm j}$ and $s_{\bm j,\bm 0}$. It is left to determine $s_{\bm l,\bm k}$ for $\bm k\neq \bm 0$. For this purpose, we choose a second set of permutations, $\bm\pi^{*(\bm k)}$, for all $\bm k\in\mbb Z_d^n\backslash\{\bm 0\}$, such that
\begin{equation}
    \pi_i^{*(\bm k)} = \left\{\begin{aligned}
        &\mathds 1,&&\quad\text{if~}k_i = 0,\\
        &(0,h_i),&&\quad\text{if~}k_i \neq 0.
    \end{aligned}\right.
\end{equation}
Obviously, $\bm\pi^{*(\bm k)}(\bm 0)=\bm h$. If we look at the outcome probability in the experiment of $X_{\bm\pi^{*(\bm k)}(\bm 0)}$,
\begin{equation}
    \Pr(\bm\pi^{*(\bm k)}(\bm t); X_{\bm\pi^{*(\bm k)}}) = c_{\bm t} + s_{\bm\pi^{*(\bm k)}(\bm t),\bm k} + O(\varepsilon^2),\quad\forall \bm t\neq \bm 0.
\end{equation}
Since we have determined all $c_{\bm t}$, the above equations determine $s_{\bm l,\bm k}$ for all $\bm l\neq\bm k,~\bm k\neq \bm 0$. Now that we have fixed all the model parameters, we conclude that there are indeed no more than $2^n-1$ gauge degrees of freedom.
We have also obtained a concrete experimental design using $(d-1)^n + d^n - 1\le2d^n$ circuits.
This completes the proof for Proposition~\ref{prop:quditSPAM}.
Note that there are $d^{2n}-2^n$ learnable degrees of freedom, and each circuit establishes at most $d^n-1$ independent linear equations. This means we need at least $d^n+1$ circuits to span all the parameters. Therefore, our number of circuits is close to the theoretical limit by a factor no more than $2$. In practice, any design that can establish a linear equation systems with column rank $d^{2n}-2^n$ can be used to extract all learnable degrees of freedom.
\noindent Examples of experiment design for a few different $n,d$ is given in Table.~\ref{tab:example}. 
\begin{table}[!htp]
    \centering
    \begin{tabular}{|c|c|c|}
       \hline
       &Num& Gates\\
       \hline
       $d=2,~n=1$ & 2 & $I,X$ \\
       \hline
       $d=3,~n=1$ & 4 & $I,X_{12},X_{01},X_{02}$\\
       \hline
       $d=4,~n=1$ & 6 & $I,X_{12},X_{13},X_{01},X_{02},X_{03}$\\
       \hline
       $d=2,~n=2$ & 4 & $I,XI,IX,XX$\\
       \hline
       \makecell{$d=3,~n=2$} & \makecell{$12$} & \makecell{$I I, X_{12} I,I X_{12},X_{12}X_{12},X_{01}I,IX_{01},X_{02}I,$\\$IX_{02},X_{01}X_{02},X_{02}X_{01},X_{01}X_{01},X_{02}X_{02}$}\\
       \hline
       $d=4,~n=2$ & 24 & \makecell{
       $II,X_{12}I,X_{13}I,IX_{12},IX_{13},X_{12}X_{12},X_{13}X_{12},X_{12}X_{13},X_{13}X_{13},$\\
       $IX_{01},IX_{02},IX_{03},X_{01}I,X_{02}I,X_{03}I,X_{01}X_{01},X_{01}X_{02},$\\
       $X_{01}X_{03},X_{02}X_{01},X_{02}X_{02},X_{02}X_{03},X_{03}X_{01},X_{03}X_{02},X_{03}X_{03}$
       }\\
       \hline
    \end{tabular}
    \caption{Example of circuit designs according to Proposition.~\ref{eq:perm_observable}.}
    \label{tab:example}
\end{table}

\end{document}